\documentclass{elsart}
\usepackage{graphicx}
\journal{Physica D}
\begin{document}

\begin{frontmatter}

\title{Exact soliton solutions of the one-dimensional 
complex Swift-Hohenberg equation}

\author[a,c]{Ken-ichi Maruno},
\ead{maruno@riam.kyushu-u.ac.jp}
\author[b]{Adrian Ankiewicz} and
\author[c]{Nail Akhmediev}

\address[a]{Research Institute for Applied Mechanics, 
Kyushu University,\\ Kasuga, Fukuoka, 816-8580, Japan}

\address[b]{Applied Photonics Group, Research School of Physical
 Sciences and Engineering,\\ 
The Australian National University, Canberra ACT 0200, Australia}

\address[c]{
Optical Sciences Centre, Research School of Physical
 Sciences and Engineering,\\ 
The Australian National University, Canberra ACT 0200, 
Australia
}

\date{\today}


\begin{abstract}
Using Painlev\'e analysis, the Hirota multi-linear method and 
a direct ansatz
technique, we study analytic solutions of the (1+1)-dimensional complex
cubic and quintic Swift-Hohenberg equations. We consider both
standard and
generalized versions of these equations. We have found that a number of
exact solutions exist to each of these equations, provided that
the coefficients
are constrained by certain relations. The set of solutions include particular
types of solitary wave solutions, hole (dark soliton) solutions and
periodic
solutions in terms of elliptic Jacobi functions and the Weierstrass $\wp$
function. Although these solutions represent only a small subset 
of the large
variety of possible solutions admitted by the complex cubic and quintic
Swift-Hohenberg equations, those presented here are the first examples of
exact analytic solutions found thus far.
\end{abstract}

\end{frontmatter}

\section{Introduction}
Complicated pattern-forming dissipative systems can be described by the
Swift-Hohenberg (S-H) equation \cite{CH93,buseta}. 
Classic examples are the
Rayleigh-B\'enard problem of convection in a horizontal fluid layer in
the gravitational field \cite{SH77,Trib}, Taylor-Couette flow \cite{HS92},
some chemical reactions \cite{Staliunas98} and large-scale flows and 
spiral core instabilities \cite{Aranson1}. 
Examples in optics include
synchronously-pumped optical parametric oscillators \cite{r4},
three-level broad-area cascade lasers \cite{Garcia} and large aspect-ratio
lasers \cite{Lega94,Weiss,Aranson97,Aranson2002}.
In the context of large-aperture lasers with small 
detuning between the atomic
and cavity frequencies, the complex cubic Swift-Hohenberg (CCSH) equation was
derived
asymptotically \cite{Staliunas99,Staliunas2001,Lega95}. 
This equation is also believed to be relevant 
for oscillatory convection in binary fluids.
The appearance of spatial patterns is the most remarkable feature of the
solutions.

Despite their visual complexity, spatial patterns are actually formed
from a certain number of ordered or chaotic combinations in space of some
simpler localized structures. It is known that the S-H equation admits
the existence of transverse localized structures and phase domains
\cite{Staliunas99}. These coherent structures can be considered as bright
or dark solitons. Consequently, these simple
localized structures and their stability are of great interest in
the study of any pattern-forming system. If we keep these visual
properties of patterns in mind, we can split the problem into
several steps.
Firstly, we should study the simplest stationary objects (localized
structures).
Secondly, we have to investigate their stability. As a last step, we
should study their interaction and the combination structures. 
The distinction between stationary and moving patterns 
is defined by the nature of the
interaction forces between the localized structures. The above program
cannot be carried out in one move. Moreover, in studying stationary localized
structures, we have to start with the simplest cases, {\it viz.}
(1+1)-dimensional structures. This sequential step-by-step approach allows us
to avoid any possible confusion inherent in trying to explain a complicated
structure in one step.

Whichever part of the above program we consider, it is clear that it can
mainly be done only using computer simulations. This has been done in the
majority
of papers published so far. We will analyse only (1+1)-dimensional bright
and dark solitons, and leave aside, for the moment, 
the question of their stability and the interaction between them. 
Moreover, our task here is to obtain some analytic results.
The latter point will further restrict the class of solutions which are
of interest in our investigation. Nonetheless, this is a step of paramount
importance because, until now, no analytic solutions have been known.

In many respects, localized structures of the S-H equation are similar to
those observed in systems described by the previously-studied
complex Ginzburg-Landau equation(CGLE).
We recall that some analytic solutions for the
cubic and quintic
CGLE are known \cite{PereiraSt,vSH92,BN85,AAbook}. 
For the cubic (1+1)-D equation, 
analytic solutions describe all possible bright and dark
soliton solutions. In contrast, in the case of the quintic equation, the
analytic solutions of the CGLE represent only a small subclass of its
soliton solutions. Moreover, the stable soliton solutions are located
outside of this subclass \cite{AAbook}. 
Therefore, practically useful results 
can only be obtained numerically.

We note that, apart from some exceptions, the CGLE generally has only
isolated solutions \cite{AAbook,Akhmediev} i.e. 
they are fixed for any particular set
of equation parameters. This property is fundamental for the whole set of
localized solutions of the CGLE. The existence of isolated solutions is one
of the basic features of dissipative systems in general. The qualitative
physical foundations of this property are given in \cite{Springer}. 
This property is one of the reasons 
why the above program is possible at all. 
Like the CGLE, the S-H equation models dissipative systems, 
and we expect that it will have this property. 
Indeed, preliminary numerical simulations 
support this conjecture.

The main difference between the S-H equation \cite{SB98} and the
CGLE lies in its more involved diffraction term. 
The latter is important in
describing more detailed features of an actual physical problem. 
However, these complications prevent us from easily analyzing 
the solutions.
In fact, it was not clear that such solutions could exist at all
\cite{r4,Weiss}. In this work, we study the quintic complex S-H
equation in 1D and report various new exact solutions.

Before going into further details, we should distinguish between the real
Swift-Hohenberg(RSH) equation and the CCSH equation. The former can be
considered as a particular case of the latter. Moreover, the RSH equation
is a phenomenological model which cannot be rigorously derived from the
original equations \cite{SH77}. In contrast, the CCSH equation is
derived asymptotically and is rigorous in the long-wavelength limit. In
standard notations, the CCSH equation can be written as
\begin{equation}
\psi_t=r\psi -(1+ic)|\psi |^2\psi+ia\Delta\psi-d(\Lambda+\Delta)^2\psi\,,
\end{equation}
where $r$ is the control parameter and $a$ characterizes the diffraction
properties of the active medium. In the limit of $\Lambda \to 0$,
i.e., in the long-wavelength limit, the differential
nonlinearities formally have higher orders and therefore can be
dropped. In this case the wave-number-selecting term,
$(\Lambda+\Delta)^2 \psi$, is just a small correction to the diffraction
term $ia\Delta \psi$. Thus the CCSH equation can be treated as a perturbed
cubic CGLE. For us the main interest here is (1+1) dimensional case when
$\Delta$ has only $x$-derivatives.

In physical problems, the quintic nonlinearity can be of equal or even
higher importance to the cubic one \cite{Moores93} as it is responsible
for stability of localized solutions.
Sakaguchi and Brand made a numerical investigation
of the complex quintic Swift-Hohenberg (CQSH) equation
\begin{equation}
\psi_t=a\psi +b|\psi |^2\psi -c|\psi|^4\psi-
d(1+\partial_{xx})^2\psi+if\partial_{xx}\psi\,, \label{cqsh}
\end{equation}
where all coefficients $a=a_1+ia_2,b=b_1+ib_2,c=c_1+ic_2$, and the order
parameter $\psi$ are complex, but $d$ and $f$ are real \cite{SB98}.
For $a,b$ and $d$ real and $c\equiv 0$ and $f\equiv 0$, the original
Swift-Hohenberg equation, which was derived as an order parameter
equation for the onset of Rayleigh-B\'enard convection in a simple
fluid, is recovered. 
Then eq.(\ref{cqsh}) can be rewritten as
\begin{eqnarray}
&&i\psi_t+(f+2id)\psi_{xx}+id\psi_{xxxx}+(b_2-ib_1)|\psi|^2\psi
+(-c_2+ic_1)|\psi|^4\psi\nonumber\\
&&\quad \quad \quad =(-a_2+i(a_1-d))\psi\,.
\end{eqnarray}
This equation can be generalized to
\begin{equation}
i\psi_t+\beta \psi_{xx}+\gamma \psi_{xxxx}+\mu |\psi|^2\psi+\nu
|\psi|^4\psi = i\delta \psi\,,\label{SwiftH}
\end{equation}
where all coefficients $\beta=\beta_1+i\beta_2\,,\,
\gamma=\gamma_1+i\gamma_2\,,\, \mu=\mu_1+i\mu_2$ and
$\nu=\nu_1+i\nu_2$ are complex and $\delta$ is real.
Eq.(\ref{SwiftH}) is written in the form which we will refer to as
the generalized CQSH (or GCQSH) equation in the rest of this study.
Then the CCHS, real quintic Swift-Hohenberg (RQSH)
and real cubic Swift-Hohenberg (RCSH) equations are particular cases of
eq.(\ref{SwiftH}) with some of the coefficients being equal to zero.

The interpretation of the variables depends on the particular
problem. In optics, $t$ is the propagation distance or the cavity
round-trip number(treated as a continuous variable), $x$ is the
transverse variable, $\beta_1$ is the 2nd order diffraction, $\gamma_1$ is
the 4th order diffraction, $\mu_2$ is a nonlinear gain (or 2-photon
absorption
if negative) and $\delta$ represents a difference between linear gain and
loss. The
angular spectral gain is represented by the coefficients $\beta_2$ and
$\gamma_2$.

Sakaguchi and Brand observed some soliton-like structures in their
numerical work \cite{SB98}.
They also showed that the CQSH equation admits a stable hole solution.
Their results indicate that exact solutions may exist, but till now
none have been found. In this paper, 
we study the CQSH and CCSH equations
by using Painlev\'e analysis and the Hirota {\it multi-linear} method.
Our method is an extension of the modified Hirota method used earlier by
Nozaki and Bekki \cite{BE84} and the modification of the Berloff-Howard 
method \cite{BH97}.
We note that Nozaki and Bekki solved the CGLE by using the Hirota 
bilinear method,
while Berloff and Howard solved some real-coefficient non-integrable
equations using the Weiss-Tabor-Carnevale (WTC) method and 
the Hirota {\it multi-linear} method. 
In addition, we confirm that our solutions are valid by using a direct
ansatz approach. The solutions we obtain here can be considered as a
basis for further development.

The rest of the paper is organized as follows. The methodology and
analytical procedures are described in Sec.~2. Painlev\'e analysis of
the CQSH and CCSH equations is performed in Sec.~3.
Exact solutions of the generalized
CQSH and CCSH equations and particular cases of them are obtained in Sec.~4.
Finally, we summarize with our conclusions in Sec.~5.

\section{Methodology}

\subsection{Painlev\'e analysis and Hirota multi-linear method}

Any solution of an equation must be in accord with the singularity
structure of that equation. Painlev\'e analysis \cite{Ablowitz}
is a tool for investigating that structure.
This analysis can be applied both for ordinary differential
equations(ODEs)
and partial differential equations(PDEs).
The power of the Painlev\'e test lies in its easy algorithmic
implementability. The main requirement is the representation of any
possible solution in the form of a Laurent expansion in the neighborhood of
a movable singularity:
\begin{equation}\label{laurent}
 u=F^{-r}\sum_{j=0}^{\infty}u_jF^j\,,
\end{equation}
where $r$ is the leading-order exponent, $F(=0)$ is a singularity
manifold given by $F(z)$,
and $u_j$ is a set of analytic functions of $z$.

There are two necessary conditions for an ODE to pass the Painlev\'e
test:
\begin{itemize}
 \item the leading-order $r$ must be an integer, and
 \item it must be possible to solve the recursion relation for the
       coefficients $u_j(z)$ consistently to any order.
\end{itemize}

The general expansion of a non-integrable equation will fail the
Painlev\'e test at one of these two steps. Leading-order analysis
can be done by balancing the highest-order derivative in
$x$ with the strongest nonlinearity.

Weiss {\it et al.} developed the singular manifold method to
introduce the Painlev\'e property into the theory of PDEs
\cite{WTC83}. A PDE is said to possess the Painlev\'e property if
its solutions are single-valued about the movable singularity
manifold. To be more specific, if the singularity manifold is
given by $F(z_1,z_2,\cdots, z_n)=0$, then a solution of the PDE
must have an expansion of the form of eqn.(\ref{laurent}).
Substitution of this expansion into the PDE determines the
positive value of $r$ (from leading order analysis) and defines
the recursion relation for the $u_j$.

Weiss {\it et al.} truncated the expansion at the ``constant term ``
level \cite{WTC83}, i.e.,
\begin{equation}
 u=u_0F^{-r}+u_1F^{-r+1}+\cdots +u_{m-1}F^{-1}+u_m\,.
\end{equation}
Substituting back into the PDE, one obtains an over-determined system of
equations for $F$ and $u_j$. The benefit of the singular manifold
method is that this expansion for a nonlinear PDE contains a lot
of information about the PDE.

Most nonlinear non-integrable equations do not possess the Painlev\'e
property, i.e., they are not free of movable critical
singularities \cite{CT89,Conte,MCC94,AAW99}.
For some equations it is still possible to obtain
single-valued expansions by putting a constraint on the arbitrary
function in the Painlev\'e expansion. Such equations are said to be
'partially integrable' systems, as presented by Hietarinta as a
generalization of the Hirota bilinear formalism for non-integrable
systems \cite{Hieta}.
He conjectured that all completely integrable PDEs can be put
into a bilinear form. There are also non-integrable equations that can be
put into the bilinear form and then the partial integrability is
associated with the levels of integrability defined by the number of
solitons that can be combined in an $N$-soliton solution. Partial
integrability means that the equation allows a restricted number of
multisoliton solutions. Berloff and Howard suggested combining
this treatment of the partial integrability with the use of the
Painlev\'e expansion, truncated before the constant term level, as
a transform to reduce a non-integrable PDE to a multi-linear
equation \cite{BH97}. The Berloff-Howard method is a powerful tool
for solving non-integrable equations. In this section, we give an
example to show how to obtain solutions using this method.

Firstly, we consider the generalized RCSH equation
\begin{equation}
 u_t+\alpha u_{xx}+\beta u_{xxxx}+\gamma u-\delta u^3=0\,,\label{rcsh}
\end{equation}
and show how to find exact solutions of non-integrable dissipative
partial differential equations by using the Painlev\'e test and
the Hirota multi-linear method.

We take the transform truncated at the term before the constant
term:
 \begin{equation}\label{expansion}
 u=u_0F^{-r}+u_1F^{-r+1}+\cdots +u_{m-1}F^{-1}\, .
\end{equation}
Analysis of the leading order terms gives $r=2$. By
substituting this expansion (\ref{expansion})
into the RCSH equation (\ref{rcsh}) and equating the coefficients
of the highest powers of $F$ to zero, we
obtain expressions for $u_0\,, u_1$ in terms of $F$, and these
lead to the transform
\begin{equation}
 u=\frac{\sqrt{\frac{120\beta}{\delta}}F_x^2}{F^2}
-\frac{\sqrt{\frac{120\beta}{\delta}}F_{xx}}{F}=
-\sqrt{\frac{120\beta}{\delta}}\frac{d^2}{dx^2}\log F\,.
\end{equation}
This transform leads to an equation which is
quadrilinear in $F$, meaning that each term contains a product
of four functions involving $F$ and its derivatives.
\begin{eqnarray}
&&\gamma F^2 F_x^2-2FF_tF_x^2
+6\alpha F_x^4+2F^2F_t F_{xt}
-\gamma F^3F_{xx}+F^2F_t F_{xx}\nonumber\\
&&\quad \quad -12\alpha FF_x^2 F_{xx}+3\alpha F^2F_{xx}^2
-90 \beta F_x^2F_{xx}^2+90 \beta FF_{xx}^3
-F^3 F_{xxt}\nonumber\\
&&\quad \quad +4 \alpha F^2F_xF_{xxx}+120\beta F_x^3F_{xxx}
-120\beta FF_xF_{xx}F_{xxx}+10\beta F^2F_{xxx}^2 \nonumber\\
&&\quad \quad -\alpha  F^3F_{xxxx}-30 \beta F F_x^2 F_{xxxx}
+15\beta F^2 F_{xx}F_{xxxx}+6 \beta F^2 F_x F_{xxxxx}\nonumber\\
&&\quad \quad -\beta F^3 F_{xxxxxx}=0\,.\label{multircsh}
\end{eqnarray}
Substituting $F=1+e^{2kx+2\omega t}$ into
eq.(\ref{multircsh}) and
equating
the coefficients of different powers of $e$ to zero, we arrive at
the one-soliton solution:
\begin{equation}
u(x)=\frac{\pm \sqrt{\frac{30\gamma}{\delta}}e^{2kx}}
{(1+e^{2kx})^2}
=\pm \frac{1}{4}\sqrt{\frac{30\gamma}{\delta}}{\rm sech}^2(kx)\,,
\end{equation}
where
\begin{equation}
 k=\pm \frac{1}{2\sqrt{2}}\left(\frac{\gamma}{\beta}\right)^{\frac{1}{4}}\,,
\end{equation}
and we require $\alpha=-\frac{5\sqrt{\beta \gamma}}{2}$ as a constraint
on the equation parameters.
This solution is reminiscent of the $\rm sech^2$-solution,
representing an ultrashort soliton at the
minimum-dispersion wavelength, taking into account the effects of
fourth order dispersion \cite{AAbook,KH94,ABK94} of the equation
\begin{equation}
 i\frac{\partial \psi}{\partial t}
+\alpha \frac{\partial^2 \psi}{\partial x^2}
+\beta \frac{\partial^4 \psi}{\partial x^4}-\delta|\psi|^2\psi=0\,.
\end{equation}
Actually, the simple ansatz $\psi=u(x)\exp(-i\gamma\,t)$
reduces this equation to the stationary real cubic
Swift-Hohenberg equation (\ref{rcsh}).

Now we consider the generalized RQSH equation in the
form \cite{SB96}
\begin{equation}
 u_t+\alpha u_{xx}+\beta u_{xxxx}+\gamma u-\delta u^3-\eta u^5=0\, .
\label{rqsh}
\end{equation}
Analysis of the leading-order terms gives us $r=1$. By
substituting eq.(\ref{expansion}) into the RQSH equation
(\ref{rqsh}) and equating the coefficients
of the highest powers of $F$ to zero, one
obtains expressions for $u_0$ in terms of $F$ that lead to the
transform
\begin{equation}
u=\left(\frac{24\beta}{\eta}\right)^{\frac{1}{4}}\frac{F_x}{F}\,.
\end{equation}
This transform again leads to a
quadrilinear equation in $F$:
\begin{eqnarray}
&&F_{tx}F^3- F_tF_xF^2+\beta F_{xxxxx}F^3
-5\beta F_{xxxx}F_xF^2-10\beta F_{xxx}F_{xx}F^2\nonumber\\
&&\quad +20\beta F_{xxx}F_x^2F+\alpha F_{xxx}F^3
+30\beta F_{xx}^2 F_xF-60\beta F_{xx}F_x^3\nonumber\\
&&-3\alpha F_{xx}F_xF^2-2\sqrt{\frac{6\beta}{\eta}}\delta F_x^3F
+2\alpha F_x^3F+\gamma F_xF^3 =0\,.\label{multirqsh}
\end{eqnarray}
On substituting $F=1+e^{2kx+2\omega t}$ into
the quadrilinear equation (\ref{multirqsh}) and equating
the coefficients of different powers of $e$ to zero, we obtain four
algebraic
equations, but these algebraic equations don't have solutions.

We suppose a different transform:
\begin{equation}
u=\frac{G}{F}\,.
\end{equation}
This leads to an equation which is
pentalinear in $F$ and $G$:
\begin{eqnarray}
&&F^4 G-F_xF^3G-\beta F_{xxxx}F^3G+8\beta F_{xxx}F_xF^2G
-4\beta F_{xxx}F^3G_x\nonumber\\
&&\quad +6\beta F_{xx}^2F^2G
-36\beta F_{xx}F_x^2FG
+24\beta F_{xx}F_xF^2G_x - 6\beta F_{xx}F^3G_{xx}\nonumber\\
&&\quad -\alpha F_{xx}F^3G+24\beta F_x^4G
-24\beta F_x^3FG_x
+12\beta F_x^2F^2G_{xx}\nonumber\\
&&\quad +2\alpha F_x^2F^2G-4\beta F_xF^3G_{xxx}-
2\alpha F_x F^3G_x
+F^4G_t+\beta F^4G_{xxxx}\nonumber\\
&&\quad +\alpha F^4G_{xx}-\delta F^2 G^3-\gamma \eta G^5=0\,.
\label{multirqsh2}
\end{eqnarray}
Substituting $F=1+e^{2kx+2\omega t}$ and
$G=2u_0\, e^{kx+\omega t}$ into equation (\ref{multirqsh2}) and
equating the coefficients of different powers of $e$ to zero, 
we get the one-soliton solution:
\begin{equation}
u=u_0\, {\rm sech}(kx)\, ,
\end{equation}
where
\begin{eqnarray}
&&k=\pm u_0 \left(\frac{\eta}{24\beta}\right)^{\frac{1}{4}}\, ,\,
u_0^2=\frac{2}{3}\,\left(-\frac{\delta}{\eta}
\pm \frac{\sqrt{\delta^2+6\gamma \eta}}{\eta}\right)\, ,\\
&&\alpha=\frac{-24\sqrt{6\beta}\gamma-\sqrt{6\beta}\eta u_0^4}
{12\sqrt{\eta}u_0^2}
 \,.
\end{eqnarray}
Substituting $F=1+e^{2kx+2\omega t}$ and
$G=u_0(-1+e^{2kx+2\omega t})$
into equation (\ref{multirqsh2}) and
equating the coefficients of different powers of $e$ to zero, we obtain
the kink solution:
\begin{equation}
u=u_0\, {\rm tanh}(kx)\, ,
\end{equation}
where
\begin{equation}
k=\pm u_0 \left(\frac{\eta}{24\beta}\right)^{\frac{1}{4}}\, ,\,
u_0^2=-\frac{\delta}{2\eta}\pm\frac{\delta^2+4\gamma \eta}{2\eta}
\, ,\,
\alpha=\frac{3\sqrt{6\beta}\gamma+2\sqrt{6\beta}\eta u_0^4}
{3\sqrt{\eta}u_0^2}
 \,.
\end{equation}
Generally speaking,
the application of this technique to complex variable equations
is difficult because of the complex
leading order.
However, we will show that this method works in some
cases such as the complex Swift-Hohenberg equation.

\subsection{Direct ansatz method}
It is well-known that solutions of many nonlinear differential
equations can be expressed in terms of hyperbolic functions like {\it tanh}
or
 {\it sech}. This fact motivates the direct method of solution in which the
starting point is a suitable ansatz so that the p.d.e. is expressible as a
polynomial in terms of {\it tanh} or {\it sech} functions. 
Clearly this method is not as general as the Hirota method, 
since it will not work if there is no
solution of the assumed form. However, in principle, the method is more
straightforward than the Hirota method, and it is more useful in obtaining
periodic solutions (in terms of elliptic functions).

In this section, we give an outline of the direct ansatz method and
illustrate facets of its application by considering some straightforward
examples.

First, we consider the RCSH equation (\ref{rcsh})
and show how to get exact solutions of non-integrable dissipative partial
differential equations by using the direct ansatz method.
The analysis of the leading-order terms gives us $r=2$, and
so we suppose the solution $u=u_0\, {\rm sech}^2(kx)$.
We obtain the soliton solution by substituting this ansatz into the
RCSH equation. This yields an algebraic equation in terms of {\it
tanh}-functions by using formulae for hyperbolic functions (see Appendix),
and we then equate the coefficients of different powers of the {\it tanh}
function to zero.
The solution is identical to that found by using the Hirota
method. We find that the constraint needed on the equation parameters
is $\gamma\,=\,\frac{4 \alpha^2}{25\,\beta}$. Assuming this is satisfied,
the solution is:
\begin{equation}
u(x)
=\pm \alpha \, \sqrt{\frac{3}{10\beta \delta}}\,{\rm sech}^2(kx)\,,
\end{equation}
where
\begin{equation}
k^2=-\frac{\alpha}{20\beta}\,.
\end{equation}

Now we consider the RQSH equation (\ref{rqsh}).
Analysis of the leading-order terms gives us $r=1$, and so
we suppose the solution $u=u_0\, {\rm sech}(kx)$. 
Substitution of this ansatz into the RQSH equation yields 
an algebraic equation in terms of the {\it tanh}
function. Again, by equating the coefficients of different powers 
of the {\it tanh} function to zero,
we obtain the soliton solution, and note that it
is identical with the Hirota method solution.

We find that the constraint needed on the equation 
parameters is now
\begin{equation}
-\frac{9\alpha^2}{100\beta}+\gamma
+\frac{3\delta^2}{50\eta}
\pm\frac{2\sqrt{6}\alpha \delta}{25\sqrt{\beta \eta}}=0\,,
\end{equation}
With this satisfied, we have:
\begin{equation}
u=u_0\, {\rm sech}(kx)\, ,
\end{equation}
where
\begin{eqnarray}
&&k=\pm \frac{u_0 \sqrt[4]{\eta}}{\sqrt[4]{24\beta}}\, ,\,
u_0^2=-\frac{6\sqrt{\beta}\delta \pm \sqrt{6\eta}\alpha}
{5\sqrt{\beta}\eta}\, ,
\end{eqnarray}
In the same way, we obtain a kink solution
by the ansatz $u_0 {\rm tanh}(kx)$
\begin{equation}
u=u_0\, {\rm tanh}(kx)\, ,
\end{equation}
where
\begin{equation}
k=\pm \frac{u_0 \sqrt[4]{\eta}}{\sqrt[4]{24\beta}}\, ,\,
u_0^2=-\frac{\delta}{2\eta}\pm\frac{\delta^2+4\gamma \eta}{2\eta}
\, ,
\end{equation}
and the constraint needed on the equation parameters is
\begin{equation}
\alpha=\frac{3\sqrt{6\beta}\gamma+2\sqrt{6\beta}\eta u_0^4}
{3\sqrt{\eta}u_0^2}\,.
\end{equation}

It is easy to develop this method to get solutions in terms of elliptic functions.
If we substitute a Jacobi elliptic function, as an ansatz, into the
RQSH equation (\ref{rqsh}), we obtain an algebraic equation
in terms of some Jacobi elliptic function. By using the formulae in the
Appendix, and equating the coefficients of different powers of the elliptic
function to zero, we then get the elliptic solutions.

The elliptic function solutions of the RQSH equation (\ref{rqsh})
are as follows:

Jacobi cn function
\begin{eqnarray}
&&u=u_0\,{\rm cn}(kx,q)\, ,
\end{eqnarray}
where
\begin{eqnarray}
&&k=\pm
\sqrt{\frac{\alpha-2\alpha q\pm
\sqrt{(2q-1)^2\alpha^2-4\beta \gamma A}}
{2\beta A}}\,,\\
&&A=16q^2-16q+1\,,\\
&&u_0^2=-\frac{2k^2q(\alpha+10\beta (2q-1)k^2)}{\delta}\,,
\end{eqnarray}
and the constraint needed on the equation parameters is
\begin{equation}
 \eta=\frac{24\beta q^2 k^4}{u_0^4}\,.
\end{equation}

Jacobi sn function
\begin{eqnarray}
&&u=u_0\,{\rm sn}(kx,q)\, ,
\end{eqnarray}
where
\begin{eqnarray}
&&k=\pm
\sqrt{\frac{\alpha+\alpha q\pm
\sqrt{(q+1)^2\alpha^2-4\beta \gamma A}}
{2\beta A}}\,,\\
&&A=q^2+14q+1\,,\\
&&u_0^2
=\frac{2k^2q(\alpha-10\beta (q+1)k^2)}{\delta}\,,
\end{eqnarray}
and the constraint needed on the equation parameters is
\begin{equation}
\eta=\frac{24\beta q^2 k^4}{u_0^4}\,.
\end{equation}

Jacobi dn function
\begin{eqnarray}
&&u=u_0\,{\rm dn}(kx,q)\, ,
\end{eqnarray}
where
\begin{eqnarray}
&&k=\pm
\sqrt{\frac{(q-2)\alpha\pm
\sqrt{(q-2)^2\alpha^2-4\beta \gamma A}}
{2\beta A}}\,,\\
&&A=q^2-16q+16\,,\\
&&u_0^2
=\frac{2k^2(-\alpha+10(q-2)\beta k^2)}{\delta}\,,
\end{eqnarray}
and the constraint needed on the equation parameters is
\begin{equation}
 \eta=\frac{24\beta k^4}{u_0^4}\,.
\end{equation}
In principle, these Jacobi elliptic function solutions can be also
obtained from multi-linear forms, because each
Jacobi elliptic function can be expressed by theta
functions which are Hirota $\tau$-functions in multi-linear form 
\cite{Chow,WW}.
However, calculation by this method is tedious, so we do not provide
details of it here.

Next we consider an elliptic function solution of the RCSH equation.
We know from Painlev\'e analysis that the RCSH equation has a double pole.
Thus we look for elliptic functions having double poles.
We suppose
\begin{equation}
u=u_0+\wp (k x)\,.
\end{equation}
Substituting this ansatz into the RCSH equation (\ref{rcsh})
yields an algebraic equation
in terms of a Weierstrass $\wp$ function.
By using formulae from the Appendix, and
equating the coefficients of different powers of the Weierstrass $\wp$
function to zero, we obtain
the following elliptic function solution:
\begin{equation}
u=u_0+\wp (k x)\,,
\end{equation}
where
\begin{eqnarray}
&&k=\pm \left(\frac{\delta }{120\beta}\right)^{\frac{1}{4}}\,,\,
u_0=\pm \frac{\alpha}{\sqrt{30\beta \delta}}\,,\\
&&g_2=\frac{2(10\beta \gamma -\alpha^2)}{3\beta \delta}\,,\,
g_3=\pm \frac{2\sqrt{2}\alpha (\alpha^2-5\beta \gamma)}
{3\sqrt{15\beta^3\delta^3}}\,.
\end{eqnarray}
From its relation to the Weierstrass $\sigma$ function (see Appendix),
we know that the Weierstrass $\sigma$ function is
a $\tau$-function of the Hirota multi-linear form.
Thus, we can construct this solution by the Hirota multi-linear method.
However, we do not give the detail of this approach here.

We can find another elliptic function solution of the RCSH equation,
\begin{equation}
u=u_0+{\rm cn}^2(k x,q)\,,
\end{equation}
where
\begin{eqnarray}
&&k=\pm \left(\frac{\delta}{120\beta q^2}\right)^{\frac{1}{4}}\,,\\
&&q=\frac{6+19u_0+15u_0^2\pm \sqrt{A}}
{8+38u_0+60u_0^2+30u_0^3}\,,\\
&&A=-15u_0^4-30u_0^3-3u_0^2+12u_0+4\,,
\end{eqnarray}
and the constraints needed on the equation parameters are
\begin{eqnarray}
 &&\alpha=20\beta(1-(2+3u_0)q)\,,\\
&&\gamma=8\beta k^4(8-(23+30u_0)q+(23+60u_0+45u_0^2)q^2)\,.
\end{eqnarray}

We note that all above hyperbolic-function solutions 
can be derived in the limit
of $q\to 1$ of above elliptic function solutions.

\section{Painlev\'e analysis of the complex Swift-Hohenberg equation}
It is difficult to use a full expansion because the CQSH equation
possesses both a complex leading order and non-integer resonances, and it
generally has no consistency conditions.
To see this, the CQSH equation is first rewritten as the following
system of equations:
\begin{eqnarray}
&& i\psi_t+(\beta_1+i\beta_2) \psi_{xx}+(\gamma_1+i\gamma_2) \psi_{xxxx}
+(\mu_1+i\mu_2) |\psi|^2\psi\nonumber\\
&&\quad \quad +(\nu_1+i\nu_2)|\psi|^4\psi=i\delta \psi,\label{cqsh1}\\
&& -i\psi_t^*+(\beta_1-i\beta_2) \psi_{xx}^*
+(\gamma_1-i\gamma_2) \psi_{xxxx}^*
+(\mu_1-i\mu_2) |\psi|^2\psi^*\nonumber\\
&&\quad \quad +(\nu_1-i\nu_2)|\psi|^4\psi^*
=-i\delta \psi^*\,.\label{cqsh2}
\end{eqnarray}

To leading order, we set $\psi$ and $\psi^*$ as
\begin{equation}
\psi\sim \psi_0 F^{-\xi_1}, \psi^*\sim \psi_0^* F^{-\xi_2}\,,
\end{equation}
where $\xi_1$ and $\xi_2$ are the leading-order exponents.
Upon equating the exponents of the dominant balance terms in CQSH
\begin{equation}
(\gamma_1+i\gamma_2) \psi_{xxxx}+(\nu_1+i\nu_2)|\psi|^4\psi\sim 0\,,
\end{equation}
we find
\begin{equation}
\xi_1+\xi_2=2\,.\label{ncondition1}
\end{equation}
To find $\xi_1$ and $\xi_2$ we must also equate the
coefficients in front of
these terms. From eq.(\ref{cqsh1}) we get
\begin{equation}
 |\psi_0|^4=-\frac{\gamma_1+i\gamma_2}{\nu_1+i\nu_2}
\xi_1(\xi_1+1)(\xi_1+2)(\xi_1+3)
\left(\frac{\partial F}{\partial x}\right)^4\,, \label{leading1}
\end{equation}
and from eq.(\ref{cqsh2})
\begin{equation}
 |\psi_0|^4=-\frac{\gamma_1-i\gamma_2}{\nu_1-i\nu_2}
\xi_2(\xi_2+1)(\xi_2+2)(\xi_2+3)\left(\frac{\partial F}{\partial
x}\right)^4\,.  \label{leading2}
\end{equation}
Eqs.(\ref{leading1}) and (\ref{leading2}) are combined to give
\begin{equation}
\frac{\gamma_1+i\gamma_2}{\nu_1+i\nu_2}
\xi_1(\xi_1+1)(\xi_1+2)(\xi_1+3)=
\frac{\gamma_1-i\gamma_2}{\nu_1-i\nu_2}
\xi_2(\xi_2+1)(\xi_2+2)(\xi_2+3)\,.  \label{ncondition2}
\end{equation}
Eqs.(\ref{ncondition1}) and (\ref{ncondition2}) are now the two equations
we need to solve
for $\xi_1$ and $\xi_2$. The result is
\begin{equation}
\xi_1=1+i \alpha\,,\, \xi_2=1-i \alpha\,,
\end{equation}
where $\alpha$ satisfies
\begin{eqnarray}
(\alpha^4-35  \alpha^2+24)\,(\gamma_1\nu_2-\gamma_2\nu_1) \,
+10\alpha\,(5 \,-\, \alpha^2 )(\gamma_1\nu_1+\gamma_2\nu_2)\,\,=\,\,0.
\label{alphaeq}
\end{eqnarray}
(To have $\alpha=0$, we need $\gamma_1
\nu_2=\gamma_2 \nu_1$ as the condition for a chirp-less solution.)
The leading-order exponents $\xi_1=1+i\alpha$ and $\xi_2=1-i\alpha$ are
not integers unless
$\alpha=0$.
This means that the CQSH equation already fails the
Painlev\'e test at the first step if $\alpha \ne\,0$.

When $\gamma_1\nu_2-\gamma_2\nu_1=0$
we can have $\alpha=0$ (or $\alpha=\sqrt{5}$). 
In fact this is the condition of the next section, 
giving the chirp-less solution.
This plain solution is not possible for the quintic CGL equation.

The expansions for $\psi$ and $\psi^*$ therefore take the form
\begin{eqnarray}
&& \psi=(\psi_0F^{-1-i\alpha}+\psi_1F^{-i\alpha}+\cdots)
\exp(iKx+i\Omega t)\,,\\
&&
\psi^*=(\psi_0^*F^{-1+i\alpha}+\psi_1F^{i\alpha}+\cdots)
\exp(-iKx-i\Omega t)\,.
\end{eqnarray}
The resonances for the CQSH equation
can be calculated from the recursion relations for the $\psi_j$'s and
$\psi_j^*$'s, but we do not include them here because they are not
important for
our purpose.

We can expect the following dependent variable transformation by the above
analysis,
\begin{equation}
 \psi=\frac{G}{F^{1+i\alpha}}\exp(i\Omega t)\,,\,
\psi^*=\frac{G^*}{F^{1-i\alpha}}\exp(-i\Omega t)\,,\label{cqshtrans}
\end{equation}
(In general, the transform should be $\psi=\frac{G}{F^{1+i\alpha}}
\exp(iKx+i\Omega t)$,
however, in our case,
this transformation also gives solutions in the next section.)

In the same way, we can obtain the dependent variable transformation
of the CCSH equation,
\begin{equation}
 \psi=\frac{G^2}{F^{2+i\alpha}}\exp(i\Omega t)\,,\,
\psi^*=\frac{{G^*}^2}{F^{2-i\alpha}}\exp(-i\Omega t)\,.\label{ccshtrans}
\end{equation}

In the next section, we show the existence of exact solutions
by using these transformations.

\section{Exact solutions of the complex Swift-Hohenberg equation}
\subsection{The complex quintic Swift-Hohenberg equation}
We consider the CQSH equation in \cite{SB98},
\begin{eqnarray}
&&i\psi_t+(f+2id)\psi_{xx}+id\psi_{xxxx}+(b_2-ib_1)|\psi|^2\psi\nonumber
+(-c_2+ic_1)|\psi|^4\psi\\
&&\quad \quad \quad =(-a_2+i(a_1-d))\psi\,,\label{cqsheq}
\end{eqnarray}
(In the numerical work of Sakaguchi and Brand, $a_2$ was 0.)

Substituting the transformation (\ref{cqshtrans}) into
the CQSH equation (\ref{cqsheq}),
we obtain a pentalinear equation
\begin{eqnarray}
&&(id-\Omega a_1+a_2-i)F^4G
+(b_2-ib_1)F^2G^2{G^*}-(c_2-ic_1)G^3{G^*}^2\nonumber\\
&&\quad-iF^3F_tG
+iF^4G_t+(2f+4id)F^2F_x^2G+24idF_x^4G
\nonumber\\
&&\quad -(2f+4id)F^3F_xG_x-24idFF_x^3G_x
-(2id+f)F^3F_{xx}G\nonumber\\
&&\quad -36idFF_x^2F_{xx}G+24idF^2F_xF_{xx}G_x
+6idF^2F_{xx}^2G
+(f+2id)F^4G_{xx}\nonumber\\
&&\quad +12idF^2F_x^2G_{xx}
-6idF^3F_{xx}G_{xx}
+8idF^2F_xF_{xxx}G-4idF^3F_{xxx}G_x\nonumber\\
&&\quad -4idF^3F_xG_{xxx}-idF^3F_{xxxx}G+idF^4G_{xxxx}=0\,.
\label{multicqsh}
\end{eqnarray}

Putting $F, G$ and $G^*$ as polynomials in terms of
$\exp(kx+\omega t)$
and substituting these functions into eq.(\ref{multicqsh}) and equating
the coefficients of different powers of $e$ to zero,
we obtain the following exact solutions. We can also obtain same
solutions by using direct ansatz method.

Bright Soliton
\begin{equation}
\psi=g\,{\rm sech}(kx)e^{i\Omega t}\, ,
\end{equation}
where
\begin{equation}
 k=\pm \sqrt{\frac{\sqrt{a_1}-\sqrt{d}}{\sqrt{d}}}\, ,
\, \Omega=-f+\frac{f\sqrt{a_1}}{\sqrt{d}}+a_2\, ,
\, |g|^2=\frac{2f(\sqrt{a_1}-\sqrt{d})}{\sqrt{d} b_2}\, .
\end{equation}
The coefficients must satisfy the following relations:
\begin{equation}
b_1=b_2\frac{2(4d-5\sqrt{da_1})}{f}\,,\,
c_1=-\frac{6b_2^2d}{f^2}\,,\,c_2=0\,.
\end{equation}

Dark Soliton
\begin{equation}
\psi=g\,{\rm tanh}(kx)
e^{i\Omega t}\, ,
\end{equation}
where
\begin{equation}
k=\pm \sqrt{\frac{fb_1+2db_2}{20db_2}}
\, ,
\, \Omega=-2fk^2+a_2\, ,\,
|g|^2=-\frac{2fk^2}{b_2}\,.
\end{equation}
The coefficients must satisfy the following relations:
\begin{equation}
a_1=d(16k^4-4k^2+1)\,,\,
c_1=-\frac{6db_2^2}{f^2}\,,\,c_2=0\,.
\end{equation}

Chirped Bright Soliton
\begin{eqnarray}
\psi=g\,{\rm sech}(kx)e^{-i\alpha \log({\rm sech}(kx))} e^{i\Omega t},
\end{eqnarray}
where
\begin{eqnarray}
&&k=\pm \sqrt{\frac{f\alpha+d(\alpha^2-1)\pm
\sqrt{d(d-a_1)A+(f\alpha+d(\alpha^2-1))^2}}{dA}}\,,\\
&&A=\alpha^4-6\alpha^2+1\,,\\
&&\Omega =k^2(f-f\alpha^2+4d\alpha +4d\alpha (1-\alpha^2)k^2))+a_2\,,\\
&&|g|^2=\frac{k^2(2d\alpha (23-7\alpha^2)k^2+6d\alpha -f(\alpha^2-2))}
{b_2}\,.
\end{eqnarray}
The coefficients must satisfy the following relations:
\begin{eqnarray}
&&b_1=\frac{k^2(3f\alpha-2d(2-\alpha^2)-2d(10-19\alpha^2+\alpha^4)k^2)}
{|g|^2}\,,\\
&&c_1=-b_2^2\frac{d(\alpha^4-35\alpha^2+24)}
{(2d\alpha (23-7\alpha^2)k^2+6d\alpha -f(\alpha^2-2))^2}\,,\\
&&c_2=-b_2^2\frac{10d\alpha(\alpha^2-5)}
{(2d\alpha (23-7\alpha^2)k^2+6d\alpha -f(\alpha^2-2))^2}\,.
\end{eqnarray}

Chirped Dark Soliton
\begin{equation}
\psi=g\,{\rm tanh}(kx)e^{-i\alpha \log({\rm sech}(k x))}e^{i\Omega t}\,,
\end{equation}
where
\begin{eqnarray}
&&k=\pm\sqrt{\frac{2}{A}-\frac{3f\alpha}{2dA}
\pm \frac{\sqrt{(4d-3f\alpha)^2-4d(d-a_1)A}}{2dA}}\,,\\
&&A=16-15\alpha^2\,,\\
&&\Omega=k^2((-2f-6d\alpha)+30d\alpha k^2)+a_2\,,\\
&&|g|^2=-\frac{k^2(2f+6d\alpha -f\alpha^2+10d\alpha(\alpha^2-8)k^2)}
{b_2}\,.\\
\end{eqnarray}
The coefficients must satisfy the following relations:
\begin{eqnarray}
&&b_1=-\frac{k^2(-4d+3f\alpha +2d\alpha^2+10d(4-5\alpha^2)k^2)}
{|g|^2}\,,\\
&&c_1=-\frac{dk^4(\alpha^4-35\alpha^2+24)}{|g|^4}\,,
\,c_2=-\frac{10d\alpha k^4(\alpha^2-5)}{|g|^4}\,.
\end{eqnarray}

Elliptic function solutions of the CQSH equation are the following:

Jacobi cn function solution
\begin{equation}
\psi=g\,{\rm cn}(kx,q)e^{i\Omega t}\, ,
\end{equation}
where
\begin{eqnarray}
&&k=\pm
\sqrt{\frac{d(1-2q)\pm
\sqrt{d(-12dq(q-1)+a_1A)}}{dA}}
\, ,\\
&&A=16q^2-16q+1\,,\\
&&\Omega=fk^2(2q-1)+a_2\,,\,|g|^2=\frac{2fqk^2}{b_2}\, .
\end{eqnarray}
The coefficients must satisfy the following relations:
\begin{equation}
b_1=-\frac{4dqk^2(1+5(2q-1)k^2)}{|g|^2}\, ,\,
c_1=-\frac{24dq^2k^4}{|g|^4}\,,\, c_2=0\, .
\end{equation}

Jacobi sn function solution
\begin{equation}
\psi=g\,{\rm sn}(kx,q)e^{i\Omega t}\, ,
\end{equation}
where
\begin{eqnarray}
&&k=\pm
\sqrt{\frac{d(q+1)\pm
\sqrt{d(-12dq+a_1A)}}{dA}}
\, ,\\
&&A=q^2+14q+1\,,\\
&&\Omega=-fk^2(q+1)+a_2\,,\,|g|^2=-\frac{2fqk^2}{b_2}\, .
\end{eqnarray}
The coefficients must satisfy the following relations:
\begin{equation}
b_1=-\frac{4dqk^2(-1+5(q+1)k^2)}{|g|^2}\, ,\,
c_1=-\frac{24dq^2k^4}{|g|^4}\,,\, c_2=0\, .
\end{equation}

Jacobi dn function solution
\begin{equation}
\psi=g\,{\rm dn}(kx,q)e^{i\Omega t}\, ,
\end{equation}
where
\begin{eqnarray}
&&k=\pm
\sqrt{\frac{d(q-2)\pm
\sqrt{d(12d(q-1)+a_1A)}}{dA}}
\, ,\\
&&A=q^2-16q+16\,,\\
&&\Omega=-fk^2(q-2)+a_2\,,\,|g|^2=\frac{2fk^2}{b_2}\, .
\end{eqnarray}
The coefficients must satisfy the following relations:
\begin{equation}
b_1=\frac{4dk^2(-1+5(q-2)k^2)}{|g|^2}\, ,\,
c_1=-\frac{24dk^4}{|g|^4}\,,\, c_2=0\, .
\end{equation}

\subsection{the generalized complex quintic Swift-Hohenberg equation}
Now we consider the generalized complex quintic Swift-Hohenberg(GCQSH)
equation
\begin{equation}
 i\psi_t+\beta \psi_{xx}+\gamma \psi_{xxxx}
+\mu |\psi|^2\psi+\nu |\psi|^4\psi=i\delta \psi\,,\label{gcqsheq}
\end{equation}
where all coefficients $\beta=\beta_1+i\beta_2,
\gamma=\gamma_1+i\gamma_2, \mu=\mu_1+i\mu_2$ and
$\nu=\nu_1+i\nu_2$ are complex and $\delta$ is real.

This equation can be easily normalized by rescaling
$t'=\mu_1t\,,\, x'=\sqrt{\frac{mu_1}{\beta_1}}x$, so that
$\beta_1$ and $\mu_1$ can be 1 if $\beta_1$ and $\mu_1$ are non zero.

Substituting the transformation (\ref{cqshtrans}) into
the GCQSH equation (\ref{gcqsheq}),
we obtain a pentalinear equation
\begin{eqnarray}
&&\nu G^3{G^*}^2
+(24+50i\alpha -35\alpha^2-
10i\alpha^3+\alpha^4)\gamma F_x^4G\nonumber\\
&&\quad +
2(-6-11i\alpha+6\alpha^2+
i\alpha^3)\gamma
(2FF_x^3G_x +3FF_x^2F_{xx}G )\nonumber\\
&&\quad +\mu F^2G^2{G^*}
- 6(-2- 3i\alpha+\alpha^2)
\gamma
(2F^2F_xF_{xx}G_x+F^2F_x^2G_{xx})\nonumber\\
&&\quad -
(-2-3i\alpha+\alpha^2)
(\beta F^2F_x^2G+
3\gamma F^2F_{xx}^2G+
4\gamma F^2F_xF_{xxx}G )\nonumber\\
&&\quad -i(-i+\alpha )
(2\beta F^3F_xG_x+
6\gamma F^3F_{xx}G_{xx}+
4\gamma F^3F_{xxx}G_x+ \nonumber\\
&&\quad 4\gamma F^3F_x G_{xxx}+
iF^3F_tG+\beta F^3F_{xx}G+
\gamma F^3F_{xxxx}G)\nonumber\\
&&\quad +(-i\delta -\Omega )
F^4G +iF^4G_t+\beta F^4G_{xx}+\gamma F^4G_{xxxx}\,.\label{multigcqsh}
\end{eqnarray}

Putting $F, G$ and $G^*$ as polynomials in terms of
$\exp(kx+\omega t)$ and
substituting these
functions into pentalinear equation (\ref{multigcqsh}) and equating
the coefficients of different powers of $e$ to zero, we get the following
solutions. We can also obtain same
solutions by using direct ansatz method.

First, we consider chirp-less($\alpha=0$) solutions.
From eq.(\ref{alphaeq}) we need $\gamma_1 \nu_2=\gamma_2 \nu_1$.

Bright Soliton
\begin{equation}
\psi=g\,{\rm sech}(kx)e^{i\Omega t}\,,
\end{equation}
where
\begin{eqnarray}
&&k=\pm \sqrt{\frac{-\beta_2\pm\sqrt{\beta_2^2+4\delta
\gamma_2}}{2\gamma_2}}
\,,\\
&&\Omega=k^2(1+\gamma_1k^2)\,,\,
|g|^2=2k^2(1+10\gamma_1k^2)\,.
\end{eqnarray}
The coefficients must satisfy the following relations:
\begin{equation}
\mu_2=\frac{2k^2(\beta_2+10\gamma_2k^2)}{|g|^2}\,,\,
\nu_1=-\frac{24\gamma_1k^4}{|g|^4}\,,\,
\nu_2=-\frac{24\gamma_2k^4}{|g|^4}\, ,
\end{equation}
and $\beta_1=\mu_1=1$ (normalized coefficients).

Dark Soliton
\begin{equation}
\psi=g\,{\rm tanh}(kx)e^{i\Omega t},
\end{equation}
where
\begin{eqnarray}
&&k=\pm \frac{1}{4}\,\sqrt{\frac{\beta_2\pm
\sqrt{\beta_2^2+16\delta \gamma_2}}{\gamma_2}}
\,,\\
&&\Omega=2k^2(8\gamma_1k^2-1)\,,\,
|g|^2=2k^2(20\gamma_1k^2-1)\,.
\end{eqnarray}
The coefficients must satisfy the following relations:
\begin{equation}
\mu_2=\frac{2k^2(20\gamma_2k^2-\beta_2)}{|g|^2}\,,\,
\nu_1=-\frac{24\gamma_1k^4}{|g|^4}\,,\,
\nu_2=-\frac{24\gamma_2k^4}{|g|^4}\, ,
\end{equation}
and $\beta_1=\mu_1=1$ (normalized coefficients).

Chirped Bright Soliton
\begin{eqnarray}
&&\psi=g\,{\rm sech}(kx)e^{-i\alpha \log({\rm sech}(kx))}
e^{i\Omega t}\,,
\end{eqnarray}
where
\begin{eqnarray}
&&k=\pm \sqrt{\frac{2\alpha +(\alpha^2-1)\beta_2
\pm \sqrt{(2\alpha +(\alpha^2-1)\beta_2)^2+4\delta A}}{2A}}\,,\\
&&A=4\alpha(\alpha^2-1)\gamma_1+(\alpha^4-6\alpha^2+1)\gamma_2\,,\\
&&\Omega=k^2((1-\alpha^2)+2\alpha \beta_2)\nonumber\\
&&\quad +k^4((1-6\alpha^2+\alpha^4)\gamma_1
+(4\alpha -4\alpha^3 )\gamma_2)\,,\\
&&|g|^2=k^2((2-\alpha^2)+3\alpha \beta_2)\nonumber\\
&&\quad +2k^4((10-19\alpha^2
+\alpha^4)\gamma_1+(23\alpha -7\alpha^3)\gamma_2)\,.
\end{eqnarray}
The coefficients must satisfy the following relations:
\begin{eqnarray}
&&\mu_2=\frac{k^2((2-\alpha^2)\beta_2-3\alpha)}{|g|^2}\nonumber\\
&&\quad +\frac{2k^4((10-19\alpha^2
+\alpha^4)\gamma_2-(23\alpha
+7\alpha^3)\gamma_1)}{|g|^2}\,,\\
&&\nu_1=-\frac{k^4((24-35\alpha^2+\alpha^4)\gamma_1
-10\alpha (\alpha^2-5)\gamma_2)}{|g|^4}\,,\\
&&\nu_2=-\frac{k^4((24-35\alpha^2+\alpha^4)\gamma_2
+10\alpha (\alpha^2-5)\gamma_1)}{|g|^4}\,,
\end{eqnarray}
and $\beta_1=\mu_1=1$ (normalized coefficients).

Chirped Dark Soliton
\begin{eqnarray}
&&\psi=g\,{\rm tanh}(kx)
e^{-i\alpha \log({\rm sech}(kx))} e^{i\Omega t}\,,
\end{eqnarray}
where
\begin{eqnarray}
&&k=\pm\sqrt{\frac{3\alpha -2\beta_2\pm
\sqrt{(3\alpha -2\beta_2)^2-4\delta A}}
{2A}}\,,\\
&&A=30\alpha \gamma_1+(15\alpha^2-16)\gamma_2\,,\\
&&\Omega=k^2(-2-3\alpha \beta_2+
(16\gamma_1-15\alpha^2\gamma_1+30\alpha \gamma_2)k^2)\,,\\
&&|g|^2\nonumber\\
&&\quad =k^2((\alpha^2-2)-3\alpha \beta_2
+10(4\gamma_1-5\alpha^2\gamma_1+8\alpha \gamma_2
-\alpha^3 \gamma_2)k^2)\,.
\end{eqnarray}
The coefficients must satisfy the following relations:
\begin{eqnarray}
&&\mu_2\nonumber\\
&&\quad =\frac{k^2((\alpha^2-2)\beta_2+3\alpha
+10(4\gamma_2-5\alpha^2 \gamma_2
-8\alpha \gamma_1
+\alpha^3 \gamma_1)k^2)}{|g|^2}\,,\\
&&\nu_1=-\frac{k^4((24-35\alpha^2+\alpha^4)\gamma_1
-10\alpha (\alpha^2-5)\gamma_2)}{|g|^4}\,,\\
&&\nu_2=-\frac{k^4((24-35\alpha^2+\alpha^4)\gamma_2
+10\alpha (\alpha^2-5)\gamma_1)}{|g|^4}\,,
\end{eqnarray}
and $\beta_1=\mu_1=1$ (normalized coefficients).

Elliptic function solutions of the GCQSH equation are the following:

Jacobi cn function solution
\begin{equation}
\psi=g\,{\rm cn}(kx,q)e^{i\Omega t}\, ,
\end{equation}
where
\begin{eqnarray}
&&k=\pm \sqrt{\frac{(1-2q)\beta_2\pm
\sqrt{(1-2q)^2\beta_2^2+4\delta \gamma_2 A}}
{2\gamma_2A}}\,,\\
&&\Omega=k^2((2q-1)+\gamma_1Ak^2)\,,\\
&& A=16q^2-16q+1\,,\\
&&|g|^2=2qk^2(1+10\gamma_1(2q-1)k^2)\,.
\end{eqnarray}
The coefficients must satisfy the following relations:
\begin{eqnarray}
&&\mu_2=\frac{2qk^2(\beta_2+10\gamma_2(2q-1)k^2)}{|g|^2}\,,\\
&&\nu_1=-\frac{24\gamma_1q^2k^4}
{|g|^4}\,,\,
\nu_2=-\frac{24\gamma_2q^2k^4}{|g|^4}\,,
\end{eqnarray}
and $\beta_1=\mu_1=1$ (normalized coefficients).

Jacobi sn function solution
\begin{equation}
\psi=g\,{\rm sn}(kx,q)e^{i\Omega t}\, ,
\end{equation}
where
\begin{eqnarray}
&&k=\pm \sqrt{\frac{(1+q)\beta_2\pm
\sqrt{(q+1)^2\beta_2^2+4\delta \gamma_2 A}}
{2\gamma_2A}}\,,\\
&&\Omega=k^2(-(q+1)+\gamma_1Ak^2)\,,\\
&& A=q^2+14q+1\,,\\
&&|g|^2=2qk^2(-1+10\gamma_1(q+1)k^2)\,.
\end{eqnarray}
The coefficients must satisfy the following relations:
\begin{eqnarray}
&&\mu_2=\frac{2qk^2(-\beta_2+10\gamma_2(q+1)k^2)}
{|g|^2}\,,\\
&&\nu_1=-\frac{24\gamma_1q^2k^4}{|g|^4}\,,\,
\nu_2=-\frac{24\gamma_2q^2k^4}{|g|^4}\,,
\end{eqnarray}
and $\beta_1=\mu_1=1$ (normalized coefficients).

Jacobi dn function solution
\begin{equation}
\psi=g\,{\rm dn}(kx,q)e^{i\Omega t}\, ,
\end{equation}
where
\begin{eqnarray}
&&k=\pm \sqrt{\frac{(q-2)\beta_2\pm
\sqrt{(q-2)^2\beta_2^2+4\delta \gamma_2 A}}
{2\gamma_2A}}\,,\\
&&\Omega=k^2(-(q-2)+\gamma_1Ak^2)\,,\\
&& A=q^2-16q+16\,,\\
&&|g|^2=2k^2(1-10\gamma_1(q-2)k^2)\,.
\end{eqnarray}
The coefficients must satisfy the following relations:
\begin{eqnarray}
&&\mu_2=\frac{2k^2(\beta_2-10\gamma_2(q-2)k^2)}{|g|^2}\,,\\
&&\nu_1=-\frac{24\gamma_1k^4}
{|g|^4}\,,\,
\nu_2=-\frac{24\gamma_2k^4}{|g|^4}\,,
\end{eqnarray}
and $\beta_1=\mu_1=1$ (normalized coefficients).

\subsection{The complex cubic Swift-Hohenberg equation}
Now we consider the CCSH equation
\begin{eqnarray}
&& \psi_t=(\mu+i\nu)\psi+if \psi_{xx}-d
(\psi+2\psi_{xx}+\psi_{xxxx})\nonumber\\
&&\quad \quad -(\delta+i\gamma)|\psi|^2\psi\,.\label{ccsheq}
\end{eqnarray}
Substituting the transformation (\ref{ccshtrans}) into the CCSH equation
(\ref{ccsheq}),
we obtain a hexalinear equation
\begin{eqnarray}
&&(-i\gamma-\delta ){G}^4{G^*}^2
-d(120+154i\alpha-71\alpha^2-14i\alpha^3+\alpha^4)F_x^4G^2\nonumber\\
&&\quad +2d(24+26i\alpha -9\alpha^2-i\alpha^3)
(4FF_x^3GG_x+3FF_x^2F_{xx}G^2)\nonumber\\
&&\quad +(\alpha^2-5i\alpha-6)
(12dF^2F_x^2G_x^2+12d
(2F^2F_xF_{xx}GG_x+F^2F_x^2GG_{xx})\nonumber\\
&&\quad +(2dF^2F_x^2G^2-ifF^2F_x^2G^2
+3dF^2F_{xx}^2G^2+4dF^2F_xF_{xxx}G^2))\nonumber\\
&&\quad +(2+i\alpha)(12dF^3F_{xx}G_x^2+24dF^3F_xG_xG_{xx}
+8dF^3F_xGG_x\nonumber\\
&&\quad -4ifF^3F_xGG_x+12dF^3F_{xx}GG_{xx}
+8dF^3F_{xxx}GG_x+8dF^3F_xGG_{xxx}\nonumber\\
&&\quad +F^3F_tG^2+2dF^3F_{xx}G^2-ifF^3F_{xx}G^2
+dF^3F_{xxxx}G^2)+2ifF^4G_x^2\nonumber\\
&&\quad -(d-\mu -i\nu +i\Omega)F^4G^2-4dF^4G_x^2
-6dF^4G_{xx}^2
-8dF^4G_xG_{xxx}\nonumber\\
&&\quad -2F^4GG_t-4dF^4GG_{xx}
+2ifF^4GG_{xx}-2dF^4GG_{xxxx}\,.\label{multiccsh}
\end{eqnarray}

Putting $F, G$ and $G^*$ as polynomials in terms of
$\exp(kx+\omega t)$ and
substituting these
functions into hexalinear equation (\ref{multiccsh}) and equating
the coefficients of different powers of $e$ to zero, we get the following
solutions. We can also obtain same
solutions by using direct ansatz method.

Chirped Bright Soliton
\begin{equation}
 \psi=g\,{\rm sech}^2(kx)e^{-\alpha \log({\rm sech}(kx))}
e^{i\Omega t}\,,
\end{equation}
\begin{equation}
k=\pm \frac{1}{\sqrt{\alpha^2-10}}\,,
\, |g|^2=-(120-71\alpha^2+\alpha^4)dk^4\,,
\end{equation}
\begin{equation}
\mu=4(9+8\alpha^2)dk^4\,,\, f=-12\alpha dk^2\,,
\end{equation}
\begin{equation}
\Omega= \nu-\alpha dk^4(96-12\alpha^2)\,,\,
\gamma=\frac{(154\alpha -14\alpha^3)dk^4}{|g|^2}\,.
\end{equation}

Now we look for solutions of elliptic function.
We suppose
\begin{equation}
\psi=(g+\wp (k x))e^{i \Omega t}\,.
\end{equation}
Substitution this ansatz into the CCSH equation (\ref{ccsheq})
yields an algebraic equation
in terms of a Weierstrass $\wp$ function
by using formulas in Appendix, and
equating the coefficients of different powers of a Weierstrass
$\wp$ function to zero, we obtain the following
elliptic function solution:
\begin{equation}
 \psi=(g+\wp (k x))e^{i \Omega t}\,,
\end{equation}
where
\begin{eqnarray}
&&k=\pm \left(\frac{-\delta }{120d}\right)^{\frac{1}{4}}\,,\,
g=\pm \sqrt{-\frac{2d}{15\delta}}\,,\,\Omega=\nu\,,\\
&&g_2=\frac{4d(5\mu-3d)}{3\delta}\,,\,
g_3=\pm \frac{4(d-5\mu)}{3}\sqrt{\frac{-2d}{15\delta^3}}\,,\\
&&f=0\, ,\,\gamma =0\,.
\end{eqnarray}

\subsection{The generalized complex cubic Swift-Hohenberg equation}
We consider the generalized complex cubic Swift-Hohenberg(GCCSH) equation
\begin{equation}
 i\psi_t+\beta \psi_{xx}+\gamma \psi_{xxxx}
+\mu |\psi|^2\psi=\delta \psi\,,\label{gccsheq}
\end{equation}
where all coefficients $\beta=\beta_1+i\beta_2,
\gamma=\gamma_1+i\gamma_2, \mu=\mu_1+i\mu_2$ and $\delta=\delta_1+
i \delta_2$ are complex.

This equation can be easily normalized by rescaling
$t'=\mu_1t\,,\, x'=\sqrt{\frac{mu_1}{\beta_1}}x$, so that
$\beta_1$ and $\mu_1$ can be 1 if $\beta_1$ and $\mu_1$ are non zero.

Substituting the transformation (\ref{ccshtrans})
into the GCCSH equation (\ref{gccsheq}),
we obtain a hexalinear equation
\begin{eqnarray}
&&\mu {G^*}^2G^4+(120+154i\alpha -71{\alpha }^2-
14i\alpha^3 +\alpha^4)\gamma F_x^4G^2\nonumber\\
&&\quad +2(-24-26i\alpha+9{\alpha }^2+ i{\alpha }^3)\gamma
( 4FG{F_x}^2F_xG_x + 3FG{F_x}^2GF_{xx} )\nonumber\\
&&\quad -(-6-5i\alpha+\alpha^2)
(12\gamma F^2F_x^2G_x^2+
24\gamma F^2F_xGG_xF_{xx}\nonumber\\
&&\quad +12\gamma F^2F_x^2GG_{xx}
+\beta F^2F_x^2G^2+3\gamma F^2F_{xx}^2G^2
+4\gamma F^2F_xF_{xxx}G^2)\nonumber\\
&&\quad -i(-2i+\alpha)F^3
(12\gamma (G_x^2F_{xx}+2F_xG_xG_{xx})\nonumber\\
&&\quad +4(\beta
F_xGG_x+3\gamma F_{xx}GG_{xx}+2\gamma F_{xxx}GG_x
+2\gamma F_xGG_{xxx})\nonumber\\
&&\quad +iF_tG^2+\beta F_{xx}G^2+\gamma F_{xxxx}G^2)
+(-i\delta -\Omega)F^4G^2\nonumber\\
&&\quad +2(\beta F^4G_x^2+3\gamma F^4G_{xx}^2
+4\gamma F^4G_x G_{xxx})\nonumber\\
&&\quad +2(iF^4GG_t+\beta F^4GG_{xx}+\gamma
F^4GG_{xxxx})\,.\label{multigccsh}
\end{eqnarray}

Putting $F, G$ and $G^*$ as polynomials in terms of
$\exp(kx+\omega t)$ and
substituting these
functions into hexalinear equation (\ref{multigccsh}) and equating
the coefficients of different powers of $e$ to zero, we get the following
solutions. We can also obtain same
solutions by using direct ansatz method.

Bright Soliton
\begin{equation}
\psi=g\,{\rm sech}^2(kx)e^{i\Omega t}\,,
\end{equation}
where
\begin{eqnarray}
&&k=\pm \frac{1}{4}\,\sqrt{\frac{5\delta_2}{\beta_2}}\,,\,
\Omega=\frac{16k^2}{5}-\delta_1\,,\\
&&|g|^2=6k^2\,,
\, \mu_2=\beta_2\,,\,
\gamma_1=-\frac{4\beta_2}{25\delta_2}\,,\,
\gamma_2=-\frac{4\beta_2^2}{25\delta_2}\,.
\end{eqnarray}
and $\beta_1=\mu_1=1$\, (normalized coefficients).

Chirped Bright Soliton
\begin{eqnarray}
&&\psi=g\,{\rm sech}^2(kx)e^{-i\alpha \log({\rm sech}(kx))}
e^{i\Omega t}\,,
\end{eqnarray}
where
\begin{eqnarray}
&&k=\pm \sqrt{\frac{-2(\alpha^2+2\alpha+10)(\alpha^2-2\alpha+10)\delta_2}
{6\alpha (\alpha^4+16\alpha^2+96)\beta_1+\beta_2 A}}\,,\\
&&A=\alpha^6+10\alpha^4+8\alpha^2-640\,,\\
&&\Omega=\frac{k^2(-(\alpha^6+10\alpha^4+8\alpha^2-640)
+6\alpha \beta_2(\alpha^4+16\alpha^2+96))}
{2(\alpha^2+2\alpha+10)(\alpha^2-2\alpha+10)}-\delta_1
\,,\\
&&|g|^2=
\frac{k^2
((-\alpha^6-3\alpha^4+94\alpha^2+1200)
+4\alpha(2\alpha^4+33\alpha^2+205)\beta_2)}
{2(\alpha^2+2\alpha+10)(\alpha^2-2\alpha+10)}\,.
\end{eqnarray}
The coefficients must satisfy the following relations:
\begin{eqnarray}
&&\mu_2=\frac{k^2
((-\alpha^6-3\alpha^4+94\alpha^2+1200)\beta_2
-4\alpha(2\alpha^4+33\alpha^2+205))}
{2|g|^2(\alpha^2+2\alpha+10)(\alpha^2-2\alpha+10)}\,,\\
&&\gamma_1=\frac{(\alpha^2-10)+6\alpha \beta_2}
{2k^2(\alpha^2+2\alpha+10)(\alpha^2-2\alpha+10)}\,,\\
&&\gamma_2=\frac{(\alpha^2-10)\beta_2+6\alpha }
{2k^2(\alpha^2+2\alpha+10)(\alpha^2-2\alpha+10)}\,,
\end{eqnarray}
and $\beta_1=\mu_1=1$ (normalized coefficients).

Now we look for solutions of elliptic function.
We suppose
\begin{equation}
\psi=(g+\wp(k x))e^{i \Omega t}\,.
\end{equation}
Substitution this ansatz into the GCCSH equation (\ref{gccsheq})
yields an algebraic equation
in terms of a Weierstrass $\wp$ function
by using formulas in Appendix, and
equating the coefficients of different powers of a Weierstrass
$\wp$ function to zero, we obtain
the following elliptic function solution:
\begin{equation}
 \psi=(g+\wp(k x))e^{i \Omega t}\,,
\end{equation}
where
\begin{eqnarray}
&&k=\pm \left(\frac{-1}
{120\gamma_1}\right)^{\frac{1}{4}}\,,\,
g=\pm \frac{1}{\sqrt{-30\gamma_1}}
\,,\,\Omega=
\frac{-2+\gamma_1(3g_2-20\delta_1)}
{20\gamma_1}\,,\\
&&g_2=\frac{2\beta_2^2+20\gamma_2\delta_2}
{3\gamma_2\delta_2}\,,\,
g_3=
\pm \frac
{2\sqrt{30}(1+5\gamma_1(\Omega+\delta_1))}
{45\sqrt{-\gamma_1^3}}
\,.
\end{eqnarray}
The coefficients must satisfy the following relations:
\begin{eqnarray}
&&\mu_2=-120\gamma_2k^4\,,\,
\beta_2^2=-30\gamma_2\mu_2 g^2
\,,\\
&&
135\gamma_2^3\mu_2^3g_3^2=
-8\beta_2^2(\beta_2^2+5\gamma_2\delta_2)^2
\,,
\end{eqnarray}
and $\beta_1=\mu_1=1$ (normalized coefficients).

We find another elliptic function solution,
\begin{equation}
\psi=(g+{\rm cn}^2(k x,q))e^{i\Omega t}\,,
\end{equation}
where
\begin{eqnarray}
&&k=\pm \left(\frac{-1}
{120\gamma_1 q^2}\right)^{\frac{1}{4}}\,,\\
&&\Omega=-\delta_1-8\gamma_1
k^4(8-(23+30g)q+(23+60g+45g^2)q^2)\,,\\
&&q=\frac{6+19g+15g^2\pm \sqrt{A}}{
2(g+1)(15g^2+15g+4)}\,,\\
&&A=-15g^4-30g^3-3g^2+12g+4\,,\\
&&g=\frac{1-6q}{3q}-\frac{1}{60\gamma_1q}\,.
\end{eqnarray}
The coefficients must satisfy the following relations:
\begin{eqnarray}
&&\mu_2=-120\gamma_2q^2k^4\,,\\
&&\beta_2=20\gamma_2(1-(2+3g)q)\,,\\
&&\delta_2=-8\gamma_2 k^4(8-(23+30g)q+(23+60g+45g^2)q^2)\,,
\end{eqnarray}
and $\beta_1=\mu_1=1$ (normalized coefficients).

\section{Conclusion}
A number of pattern formation phenomena are descrived by the (2+1)
dimensional complex Swift-Hohenberg equation. Among them is an
approximation of a Maxwell-Bloch system for the transverse
dynamics of a two-level class B laser, filamentation in wide aperture
semiconducter lasers etc. (1+1) dimensional version of
this equation is one of the basic equations for modelling
temporal behaviour of laser systems. Among others, it describes, short pulse
generaion by passively mode-locked lasers with complicated spectral
filtering elements.
The knowledge of its solutions presented in any form will help to understand
better the processes in such systems. In this paper, using Painlev\'e
analysis, Hirota multi-linear method and direct ansatz technique, we studied
analytic solutions of the (1+1)-dimensional complex cubic and quintic
Swift-Hohenberg equations. We considered both, standard and generalized
versions of these equations. We have found that a number of exact solutions
exist to each of these equations provided that coefficients are bound by
special relations. The set of solutions include particular types of solitary
wave solutions, hole (dark soliton) solutions and periodic solutions in
terms of elliptic Jacobi functions and Weierstrass $\wp$ function. Clearly
these solutions represent only a small subset of large variety of possible
solutions admitted by the complex cubic and quintic Swift-Hohenberg
equations. Nevertheless, the solutions presented here are found for the
first time and they might serve as seeding solutions for a wider class of
localised structures which, no doubt, exist in these systems. We also hope
that they will be useful in further numerical analysis of various solutions
to the complex Swift-Hohenberg equation.

\section{Acknowledgements}
K.M. is grateful to H. Sakaguchi and N. Berloff for
stimulating discussions and helpful suggestions. 
K.M. was supported by a JSPS Fellowship for Young Scientists.
N.A. acknowledges financial support from the Secretar\'ia de Estado de
Educaci\'on y Universidades, Spain, Reference No. SAB2000-0197 and
support from US AROFE(grant N62649-02-1-0004).)

\appendix
\section{Formulas of hyperbolic functions and elliptic functions}

Let ${\rm Dn}={\rm dn}(k x, q),\,{\rm Sn}={\rm sn}(k x, q),\,{\rm
Cn}={\rm cn}(k x, q)$.

Jacobi elliptic functions satisfy the following two relations:
\begin{eqnarray}
&&{\rm Sn}^2+{\rm Cn}^2=1\,,\\
&&{\rm Dn}^2=1-q {\rm Sn}^2\,.
\end{eqnarray}

We have the following differential formulas:
\begin{equation}
\frac{1}{k^2}\,\frac{{\rm Dn}''(x)}
{{\rm Dn}(x)}=q(-1+2 {\rm Sn}^2)\,,
\end{equation}
and
\begin{equation}
\frac{1}{q k^4}\,
\frac{{\rm Dn}''''(x)}{{\rm Dn}(x)}
=4+q-4(2+5q){\rm Sn}^2+24q{\rm Sn}^4\,.
\end{equation}
Now the limit $q=1$ reduces this to
${\rm dn}(k x, 1)={\rm sech}(kx)=S(x),
\,{\rm sn}(k x,1)={\rm tanh}(kx)=T(x)$. Then
\begin{equation}
\frac{1}{k^2}\,\frac{S''(x)}{S(x)}= -1+2T^2\,,
\end{equation}
and
\begin{equation}
\frac{1}{k^4}\,\frac{S''''(x)}{S(x)}=5-28T^2+24T^4 \,.
\end{equation}

The other main function we need for the periodic solutions is Jacobi
sn function.
We have the following differential formulas:
\begin{equation}
\frac{1}{k^2}\,\frac{{\rm Sn}''(x)}
{{\rm Sn}(x)}=2q{\rm Sn}^2(x)-q-1\,,
\end{equation}
and
\begin{eqnarray}
\frac{1}{k^4}\,\frac{{\rm Sn}''''(x)}{{\rm Sn}(x)}&=&
1+14q+q^2
-20q(q+1){\rm Sn}^2(x)\nonumber\\
&\quad& +24q^2{\rm Sn}^4(x)\,.
\end{eqnarray}
Now the limit $q=1$ reduces this to ${\rm sn}(k x, 1)
={\rm tanh}(k x)=T(x)$. Then
\begin{equation}
\frac{1}{k^2}\,\frac{T''(x)}{T(x)}=2(-1+T^2)\,,
\end{equation}
and
\begin{equation}
\frac{1}{k^4}\,\frac{T''''(x)}{T(x)}=8(2-5T^2+3T^4) \,.
\end{equation}

We give the following formula of Jacobi cn function:
\begin{equation}
\frac{1}{k^2}\,\frac{{\rm Cn}''(x)}
{{\rm Cn}(x)}=-1+2q{\rm Sn}^2(x)\,,
\end{equation}
and
\begin{eqnarray}
\frac{1}{k^4}\,\frac{{\rm Cn}''''(x)}
{{\rm Cn}(x)}&=&1+4q
-4q(2q+5){\rm Sn}^2(x)\nonumber\\
&\quad& +24q^2{\rm Sn}^4(x)\,.
\end{eqnarray}

Finally, we give the following differential formulas of Weierstrass
$\wp$ function:
\begin{eqnarray}
&&[\wp'(z)]^2=4[\wp(z)]^2-g_2\wp(z)-g_3\,,\\
&&\wp''(z)=6[\wp(z)]^2-\frac{1}{2}g_2\,,\\
&&\wp''''(z)=12\left(10[\wp(z)]^3-\frac{3}{2}g_2\wp(z)-g_3\right)\,,
\end{eqnarray}
where $g_2$ and $g_3$ are constants.
Weierstrass $\wp$ function is connected to Weierstrass $\sigma$ function
by
\begin{equation}
 \wp(x)=-\frac{{\rm d}^2}{{\rm d}x^2}\log \sigma (x)\,.
\end{equation}
Weierstrass $\wp$ function can be expressed by Jacobi elliptic
functions:
\begin{equation}
\wp(x)=e_3+\frac{e_1-e_3}{{\rm sn}^2(k x,q)}
\,,\,k^2=e_1-e_3\,,\,q=\frac{e_2-e_3}{e_1-e_3}\,,
\end{equation}
where
\begin{equation}
e_1+e_2+e_3=0\,,\,e_1e_2+e_1e_3+e_2e_3=-\frac{g_2}{4}\,,\,e_1 e_2
e_3=\frac{g_3}{4}.
\end{equation}

\end{document}